\documentclass[ prl, superscriptaddress]{revtex4-2}

\usepackage{graphicx}
\usepackage{dcolumn}
\usepackage{bm}%
\usepackage{amsmath}
\graphicspath{ {./images/} }
\usepackage[colorlinks=true, allcolors=blue]{hyperref}
\usepackage{amsfonts}
\usepackage{physics}
\usepackage{subcaption}
\usepackage{mathtools}
\usepackage{setspace}
\usepackage{nameref}
\usepackage[font=small]{caption}
\usepackage[colorlinks=true, allcolors=blue]{hyperref}

\begin{document}
\doublespacing
\preprint{APS/123-QED}

\title{Mixed State Entanglement Via the Cauchy-Schwarz Inequality}
\author{Nishaant Jacobus}
\affiliation{Chemical Physics Theory Group, Department of Chemistry, and Center for Quantum Information and Quantum Control,
University of Toronto, Toronto, Ontario M5S 3H6, Canada}
\author{Paul Brumer}
\affiliation{Chemical Physics Theory Group, Department of Chemistry, and Center for Quantum Information and Quantum Control,
University of Toronto, Toronto, Ontario M5S 3H6, Canada}
\author{Chern Chuang}
\affiliation{Department of Chemistry and Biochemistry, University of Nevada, Las Vegas, Las Vegas, Nevada 89154, USA}
\date{\today}

\begin{abstract}
The entanglement properties of mixed states are of great importance in the study of open quantum systems and quantum information science, but commonly used entanglement measures, such as negativity, can be difficult to apply or connect to physical properties of the system. We introduce the Cauchy-Schwarz Violation (CSV) Condition, which has a simple dependence on the populations and coherences of the density operator. A sufficient condition for entanglement, it provides a more direct connection to the physical characteristics of the system such as its symmetries. We illustrate the often surprising insights gained from the CSV condition by applying it to the Jaynes-Cummings Model, the Quantum Rabi Model, and an open-system Quantum Rabi Model.
\end{abstract}
\maketitle
Since its introduction by Schrödinger nearly a century ago \cite{schrodinger_gegenwartige_1935}, quantum entanglement has been of great interest, with a central role in quantum computing and quantum information science \cite{preskill_quantum_2018, benenti_principles_2018}, a possible influence on chemical processes \cite{molina-espiritu_quantum_2015, li_entanglement_2019, whaley_quantum_2011, caruso_entanglement_2010}, and the potential to be the fundamental difference between quantum and classical mechanics \cite{karimi_classical_2015, paneru_entanglement_2020, korolkova_operational_2024-1}. 
Due to this importance, several measures have emerged attempting to quantify the entanglement of statistical mixtures of quantum states (mixed states) \cite{plenio_introduction_2006, elben_mixed-state_2020}, with one of the most popular being the  entanglement negativity $\mathcal{N}(\rho)$ \cite{peres_separability_1996, horodecki_separability_1996}. Here, $\mathcal{N}(\rho) = \sum_{\lambda_k \in\, \textrm{spec}(\rho^{PT}), \lambda_k < 0} |\lambda_k|$
where  $\textrm{spec}(\rho^{PT})$ denotes the spectrum of $\rho^{PT}$ and the $PT$ superscript denotes the partial transpose of the density operator $\rho$, given by $\langle i, j|\rho^{PT}|k,l\rangle = \langle k,j|\rho|i,l\rangle$ in any product basis. Negativity has proven successful as a quantitative measure of entanglement \cite{benabdallah_quantum_2025, brenes_bath-induced_2024, parez_entanglement_2024}, but can be challenging to compute as well as to interpret physically. That is, it is generally difficult to read off the eigenvalues of a high-dimensional operator, much less link those eigenvalues to characteristics such as populations, coherences, and symmetries. 

In this Letter we introduce an alternative characterization for entanglement based on the Cauchy-Schwarz inequality that directly depends on the populations and coherences of the system, enabling immediate connections between entanglement and physical structure. By applying this approach to the Jaynes-Cummings Model (JCM) and Quantum Rabi Model (QRM), we demonstrate the ease of use of this method, as well as the new physical insights it provides about the connection between symmetry and entanglement. The general principles of the CSV condition illustrated by these examples also apply to larger systems, where negativity is difficult to calculate either analytically or numerically.

\textit{The CSV Condition}---Because negativity measures the extent to which $\rho^{PT}$ fails to be positive semidefinite (i.e. $\mathcal{N}(\rho) = 0$ if and only if $\rho^{PT}$ is positive semidefinite), a new characterization of entanglement requires other ways to determine whether an operator is positive semidefinite. Given a Hermitian matrix $A$, it follows from the Cauchy-Schwarz inequality that if $A$ is positive semidefinite, then $A_{ii}A_{jj} \ge |A_{ij}|^2$ for any $i,j$ \cite{nitzan_chemical_2014, skorobagatko_universal_2021, cohen-tannoudji_quantum_2019}. Consequently, if one defines the quantity $\mathcal{S}(\rho) = \sum_{i > j} \max\{|\rho_{ij}^{PT}|^2 - \rho_{ii}^{PT}\rho_{jj}^{PT}, 0 \}$, then a nonzero $\mathcal{S}$ indicates that $\rho^{PT}$ is not positive semidefinite, and hence $\mathcal{N}(\rho) > 0$. We refer to this distinction between zero and nonzero $\mathcal{S}$ as the Cauchy-Schwarz Violation (CSV) Condition. Because it only involves two populations and one coherence at a time, $\mathcal{S}$ is a highly ``local" detection of entanglement, in contrast to negativity, which is inherently ``global", involving diagonalization of the entire density operator. The locality of $\mathcal{S}$ makes it easier to study both analytically and numerically, and will enable connections, as discussed below, between $\mathcal{S}$ and other physical properties of quantum systems such as symmetries. 
Additionally, one would expect that a larger $\mathcal{S}$ should generally correspond to a larger magnitude of  $\mathcal{N}$. At present we do not make this latter correspondence mathematically precise, but we will show that it does hold for several examples throughout the work. 

Before applying $\mathcal{S}$ to study specific systems, consider what information is discarded by considering $\mathcal{S}$ instead of $\mathcal{N}$. By Sylvester's Criterion, an $n \times n$ Hermitian matrix $A$ is positive semidefinite if and only if all principal minors (determinants of submatrices obtained from $A$ by deleting a set of rows and the corresponding set of columns) are nonnegative \cite{meyer_matrix_2000}. In the context of the density operator $\rho$, each minor serves as a ``source" of negativity, with the minors of larger submatrices representing more ``delocalized" entanglement involving many populations and coherences. Local sources of entanglement arise from the principal minors associated with the smallest submatrices. The principal minors from $1 \times 1$ submatrices of $\rho^{PT}$ are nonnegative if and only if $\rho_{ii} \ge 0$, which is automatically satisfied. Therefore, the most local source of negativity comes with the principal minors of $2 \times 2$ submatrices. For these, a nonnegative principal minor occurs if and only if $\rho_{ii}^{PT}\rho_{jj}^{PT} \ge |\rho_{ij}^{PT}|^2$, precisely the inequality used to construct the CSV condition. So, from the perspective of the Sylvester Criterion, $\mathcal{S}(\rho)$ measures the most local source of negativity, providing a sufficient but not necessary test for entanglement. CSV is a ``lowest order" approximation of the negativity, and therefore cannot always detect nonzero negativity because it ignores ``higher order" sources of entanglement arising from the larger submatrices. Nonetheless, accounting for only local negativity sources through $\mathcal{S}$ is often sufficient to qualitatively understand the negativity properties of a system.

\textit{Application: Low-Temperature Jaynes-Cummings and Quantum Rabi Models}---To illustrate the general principles of the CSV approach, we compare the entanglement properties of the JCM and QRM. These are two related models that describe a two-level system (TLS) interacting with a single bosonic mode \cite{mirzaee_atomfield_2015, akella_dynamics_2022, mandal_entanglement_2024, boukobza_entropy_2005, fan_quantum_2020, yang_characterizing_2023, bina_entanglement_2014, shi_entanglement_2022, ying_ground-state_2015, liu_quantum_2023, rossatto_spectral_2017, chuang_parametric_2023}. Comparing these models will both reveal the correlation between the magnitudes of $\mathcal{S}$ and $\mathcal{N}$ and demonstrate how the CSV condition provides a direct connection between symmetry and entanglement. The QRM is given by the Hamiltonian  $H_{QRM} = \frac{\Delta}{2}\sigma_z + a^\dagger a + \lambda\sigma_x (a + a^\dagger)$,
with $\sigma_z, \sigma_x$ the Pauli matrices and $a^\dagger$ ($a$) the creation (annihilation) operator for a bosonic mode.
One can then obtain the Hamiltonian for the JCM via the rotating wave approximation, which discards the counterrotating terms in the interaction to yield $ H_{JCM} = \frac{\Delta}{2}\sigma_z + \lambda(\sigma^-a^\dagger + \sigma^+a) +  a^\dagger a$
with $\sigma^{\pm}$ the raising/lowering operators for a TLS. Due to the lack of counterrotating terms, the JCM conserves the excitation number $N = (\sigma_z + 1) + a^\dagger a$. On the other hand, the QRM only has a $\mathbb{Z}_2$ symmetry, preserving the parity $P^{\mathbb{Z}_2} = -\sigma_z(-1)^{a^\dagger a}$ of the total excitation count.

\begin{figure}
    \centering
\includegraphics[width=\linewidth]{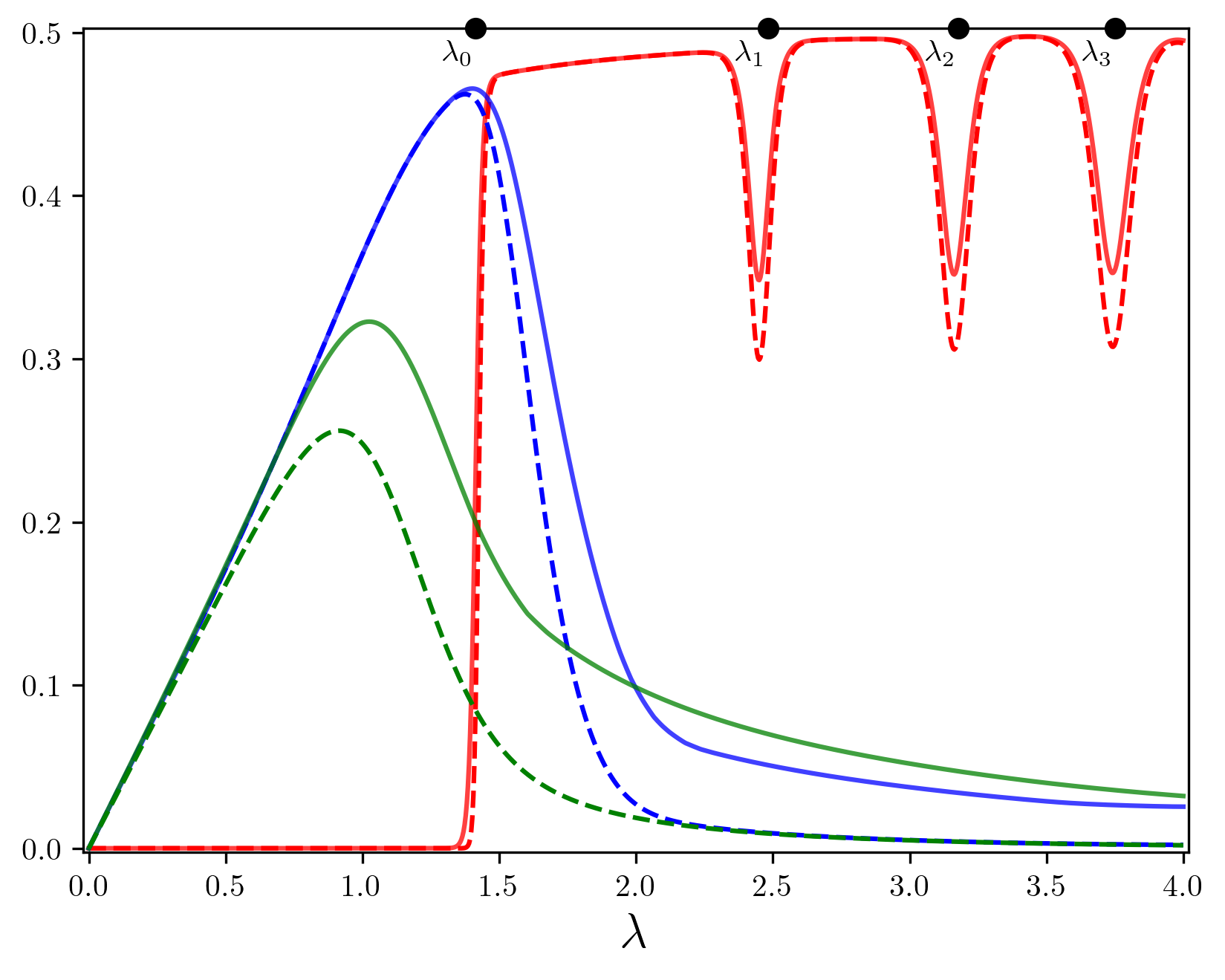}
        \caption{Numerically calculated $\sqrt{\mathcal{S}}$ (solid lines) and $\mathcal{N}$ (dashed lines) for the thermal states of $H_{JCM}$ (red), $H_{QRM}$ (blue), and $H_\varepsilon$ for $\varepsilon = 0.1\Delta$ (green), with the JCM degeneracies $\lambda_n$ marked at the top of the figure. For all calculations, $\Delta = 2, \beta = 90$ with other details of the numerical calculations in the Supplemental Material.}
        \label{JC_Rabi_Cross_Section}
\end{figure}

Consider first the low-temperature thermal states of the JCM. This model can be solved exactly using the excitation number symmetry (see Supplemental Material for details). Defining the ``critical" (meaning associated with degeneracy) coupling values $\lambda_0 = \sqrt{\Delta}$, and $\lambda_n =\sqrt{2n + 1 + \sqrt{(2n + 1)^2 + (\Delta - 1)^2}}
$ for $n \ge 1$, one finds that when $|\lambda| \in (\lambda_{n-1}, \lambda_n)$ (interpreting $\lambda_{-1}$ as 0), there is a unique ground state $|G\rangle$ with $n$ excitations, while for $|\lambda| = \lambda_n$, the ground state is two-fold degenerate, with one state $|G_1\rangle$ having $n$ excitations and the other $|G_2\rangle$ having $n + 1$. Consequently, in the zero-temperature limit, the thermal state $\rho_{Th} = e^{-\beta H}/Z$ can be written as $\rho_{Th} = |G\rangle\langle G|$ when $\lambda$ is not critical, and $\rho_{Th} = \frac{1}{2}\left(|G_1\rangle\langle G_1| + |G_2\rangle\langle G_2| \right)$ otherwise. When $\lambda_{n-1} < |\lambda| < \lambda_{n}$, we find $\mathcal{S}(\rho_{Th}) =\frac{\lambda^2n}{(\Delta - 1)^2 + 4\lambda^2 n}$ and $\mathcal{N}(\rho_{Th}) = \sqrt{\mathcal{S}(\rho_{Th})}$, making the numerical correspondence between $\mathcal{S}$ and $\mathcal{N}$ precise in this case. In particular, $\mathcal{S}$ ($\mathcal{N}$) remains $0$ until $|\lambda| \ge  \lambda_0$, then quickly asymptotes to $1/4$ ($1/2$) with increasing $|\lambda|$. On the other hand, when $|\lambda| = \lambda_n$, the expressions for $\mathcal{S}(\rho_{Th})$ and $\mathcal{N}(\rho_{Th})$ are more complicated and no longer have a simple relation. However, the magnitudes of $\mathcal{S}$ and $\mathcal{N}$ are still qualitatively correlated, as both decrease relative to the non-critical $\lambda$, with $\mathcal{S}$ now bounded by $1/8$ and $\mathcal{N}$ by $1/3$. This is further verified numerically by calculating the thermal steady state of the JCM and its corresponding $\sqrt{\mathcal{S}}$ and $\mathcal{N}$ at low but nonzero temperature; the results are shown in red in Fig. \ref{JC_Rabi_Cross_Section}. We see interesting entanglement behavior as evident in $\sqrt{\mathcal{S}}$ and $\mathcal{N}$: both are negligible for $|\lambda| < \lambda_0$ and decrease whenever $\lambda = \lambda_n$ for some $n$, with the dips broadened by the nonzero temperature.

This dependence of $\mathcal{S}$, and hence of $\mathcal{N}$, on $\lambda$ is a consequence of the interplay of the CSV condition and the excitation number symmetry in the JCM. When $\lambda \neq \lambda_m$ for any $m$, the ground state is non-degenerate and has a definite excitation number $n$; consequently, the thermal density operator only has coherences between states with excitation number $n$. However, because the excitation number depends on both the TLS and the bosonic degree of freedom, it can be changed by a partial transpose: a coherence between states with excitation number $n$ can become a coherence between a state with excitation number $n - 1$ and a state with excitation number $n + 1$. Since only states with excitation number $n$ are populated in the non-degenerate case, there is no population available to ``support" these partial transposed coherences (in the sense of maintaining the Cauchy-Schwarz inequality $\rho_{ii}^{PT}\rho_{jj}^{PT} \ge |\rho_{ij}^{PT}|^2$); this results in a large $\mathcal{S}$. For example, when $n = 1$ the ground state has coherences between $|\uparrow, 0\rangle$ and $|\downarrow, 1 \rangle$ (where the first entry of the ket denotes the spin state, and the second denotes the number of bosonic excitations). Under partial transpose, this becomes a coherence between $|\downarrow, 0\rangle$ and $|\uparrow,1\rangle$, but neither of these states are populated because their excitation numbers differ from 1. The only exception to the above discussion is when $|\lambda| \in (0, \lambda_0)$. In this case, the ground state has zero excitations, but the zero-excitation subspace is one-dimensional, hence in this case the thermal state cannot have any coherence to partial transpose, and instead we are left with $\mathcal{S} = \mathcal{N} = 0$.

 In contrast, as $\lambda$ approaches $\lambda_n$, the ground state becomes degenerate and the low-temperature thermal state populates two states, one with $n$ excitations and the other with $n + 1$, resulting in coherences between states with $n$ excitations and between states with $n + 1$ excitations. Thus, there are now twice as many coherences, but each only has half the magnitude since each state only has half population at the ground state degeneracy. Since $\mathcal{S}$ scales as a sum of squares of the coherences, it should therefore be approximately halved compared to the non-degenerate case, corresponding to the decrease at $|\lambda| = \lambda_n$ we found above. Altogether, we see that the behavior of $\mathcal{S}$ is closely tied to the conservation of excitation number in the JCM, establishing a connection between symmetry and entanglement. 

Given this link between symmetry and entanglement, one would expect the QRM to display entanglement properties different from those of the JCM because it only has a $\mathbb{Z}_2$ symmetry. Examine then how this difference can be predicted from the CSV condition. Using the $\mathbb{Z}_2$ symmetry, the eigenstates of $H_{QRM}$ can be labeled as even (if they are $+1$ eigenstates of $P^{\mathbb{Z}_2}$) or odd ($-1$ eigenstates of $P^{\mathbb{Z}_2}$). The even (odd) states consist of superpositions of states with an even (odd) number of total excitations. Because the symmetry involves information about both subsystems, the partial transpose can move coherences between states with different symmetries, transforming certain coherences between even states (``even-even coherences") into coherences between odd states (``odd-odd coherences"), and vice versa (for example, $|\uparrow, 0\rangle\langle \downarrow, 1| \to |\downarrow,0\rangle \langle \uparrow, 1|$).

As we did for the JCM, consider low-temperature thermal states of the QRM, and examine how entanglement for these states depends on the coupling strength $\lambda$. When $\lambda \ll \Delta$, the TLS and the bosonic mode are only weakly coupled and the ground state $|G\rangle$ is dominated by the uncoupled ground state $|\downarrow, 0\rangle$, weakly mixed with other even states. At low temperatures, the thermal state is approximately  $|G\rangle\langle G|$, which has weak even-even coherences due to the mixing of the ground state, but no odd population. Consequently, upon partial transposing even-even coherences into odd-odd ones, there is no population to support them, so $\mathcal{S}$ will be small but nonzero. 

In the intermediate coupling regime $\lambda \approx \Delta$, the larger $\lambda$ leads to stronger mixing of the ground state, resulting in larger even-even coherences. However, since $\Delta$ is still comparable to $\lambda$, one expects it to keep the energy gap between the ground state (even) and the first excited state (odd) large enough so that only the ground state has a significant population at low temperatures. This combination of large even-even coherences and a negligible odd population ensures that partial transposition leads to a large $\mathcal{S}$. 

In the strong-coupling regime $\lambda \gg \Delta$, one can view $\frac{\Delta}{2}\sigma_z$ as a perturbation to $H_0 = \lambda\sigma_x(a + a^\dagger) + a^\dagger a$. 
By working in the $x$-polarization basis and viewing the bosons as shifted harmonic oscillators, it is straightforward to see that all energy levels of $H_0$ are doubly degenerate, with one even state and one odd state per level. Since $\Delta \ll \lambda$, the $\frac{\Delta}{2}\sigma_z$ perturbation only slightly splits the degenerate ground state
, so that at low but nonzero temperature, both of the unperturbed ground states are significantly populated. This ensures that the thermal state has both significant even and odd population. Consequently, although there are large even-even \textit{and} odd-odd coherences due to the strong coupling $\lambda$, these coherences can be supported upon partial transposition because there is a population of each type of symmetry. Thus, surprisingly, $\mathcal{S}$ is expected to be suppressed in the strong-coupling regime. 

Therefore, from symmetry considerations alone, we expect $\mathcal{S}$ to depend non-monotonically on the coupling strength $\lambda$, suggesting that the negativity also behaves non-monotonically. This non-monotonicity and correlation between $\mathcal{S}$ and $\mathcal{N}$ are confirmed by numerical calculations, with an example shown in blue in Fig. \ref{JC_Rabi_Cross_Section}. While the precise relation $\mathcal{N} = \sqrt{\mathcal{S}}$ no longer holds, we see that $\mathcal{N}$ and $\sqrt{\mathcal{S}}$ are still quantitatively similar. As further verification, one can check that the magnitudes of the populations and coherences of the numerically calculated thermal states match the qualitative predictions given above based on symmetry (details in Supplemental Material). 

From Fig. \ref{JC_Rabi_Cross_Section}, one sees that the different symmetries of the JCM and QRM lead to remarkably different entanglement behavior. Surprisingly, in the strong-coupling regime the entanglement in the QRM \textit{decreases} with increasing coupling strength $\lambda$, unlike the JCM where it plateaus. On the other hand, in the weak-coupling regime, the QRM exhibits nonzero entanglement so long as the coupling $\lambda$ is nonzero, whereas in the JCM, there is no entanglement until $|\lambda|$ passes a threshold value. These disparate behaviors in the weak-coupling regime are remarkable because the only difference between the QRM and JCM is the presence of counterrotating terms, which one expects to have a negligible effect in the weak-coupling regime. Similar to what has been seen in the studies of entanglement of coupled TLS \cite{scala_dissipation_2008, wang_steady-state_2019-1}, we see here that the counterrotating terms in fact have a significant effect on entanglement even in the weak-coupling regime because they alter the symmetry of the system. The CSV condition provides the key insight that the counterintuitive differences in the thermal entanglement properties of the JCM and QRM, both in the weak- and strong-coupling regimes, is a consequence of the differing symmetries of the two models.

Finally, as additional evidence of the connection between symmetry and entanglement, we can further reduce the symmetry by considering a modified QRM Hamiltonian $H_\varepsilon = \frac{\Delta}{2}\sigma_z + \varepsilon \sigma_x  + \lambda\sigma_x(a + a^\dagger) + a^\dagger a$. According to the CSV condition, the buildup of negativity in the intermediate coupling regime of the QRM is due to the $\mathbb{Z}_2$ symmetry forbidding odd population, which induces a large $\mathcal{S}$. Consequently, with the $\mathbb{Z}_2$ symmetry destroyed when $\varepsilon \neq 0$, we should expect the buildup of $\mathcal{S}$ and $\mathcal{N}$ to be suppressed. Indeed, numerical calculations show that in the intermediate-coupling regime, both  $\mathcal{S}$ and $\mathcal{N}$ are strongly suppressed in this modified system, with a large $\varepsilon$ leading to a greater reduction of entanglement. For example, the green curves in Figure \ref{JC_Rabi_Cross_Section} show that even for a small perturbation of $\varepsilon = 0.1\Delta$, the maximal negativity is approximately halved. Once again, the CSV condition provides an easy way to understand how modifying a closed-system Hamiltonian will alter its entanglement properties. 

\textit{Rabi Model with Symmetry-Respecting Baths}---So far, we have applied the CSV condition to study how the symmetries of the closed-system Hamiltonian impact entanglement. However, there has been growing interest in open quantum systems where the bath also respects the symmetry of the system, as this can lead to multiple steady states or a loss of ergodicity \cite{kawabata_lieb-schultz-mattis_2024, kawabata_symmetry_2023, albert_symmetries_2014, buca_note_2012, sa_symmetry_2023, zhou_reviving_2025}. We briefly explore how the CSV condition can be helpful in studying these scenarios. 

 Consider a QRM that interacts with two independent bosonic baths via the interaction $H_{sb} = \gamma_eP_eTP_e\sum_{k} g_{k} (b_{k}^\dagger + b_{k}) + \gamma_o P_oTP_o\sum_{l} g_{l} (c_{l}^\dagger + c_{l})$, where $P_e = (1 +  P^{\mathbb{Z}_2})/2$ and $P_o = (1 -  P^{\mathbb{Z}_2})/2$ are the even and odd projectors and $T = \sigma_x(a^\dagger + a)$. Here, $a^{(\dagger)}$ is the annihilation (creation) operator of the bosonic mode of the QRM, while $b^{(\dagger)}$ and $c^{(\dagger)}$ correspond to the modes of the bath. $\gamma_e, \gamma_o, g_{k}$ and $g_{l}$ determine the coupling strength with the bath modes, and the summations $k, l$ are taken over the bosonic modes of each bath. The projection operators ensure that the first bath (with the $b^{(\dagger)}$ operators) thermalizes only the even subspace, while the second bath (with the $c^{(\dagger)}$ operators) thermalizes only the odd subspace. Therefore, from now on the first bath is referred to as the even bath, and the second as the odd bath. 
 \begin{figure}
    \centering
\includegraphics[width=\linewidth]{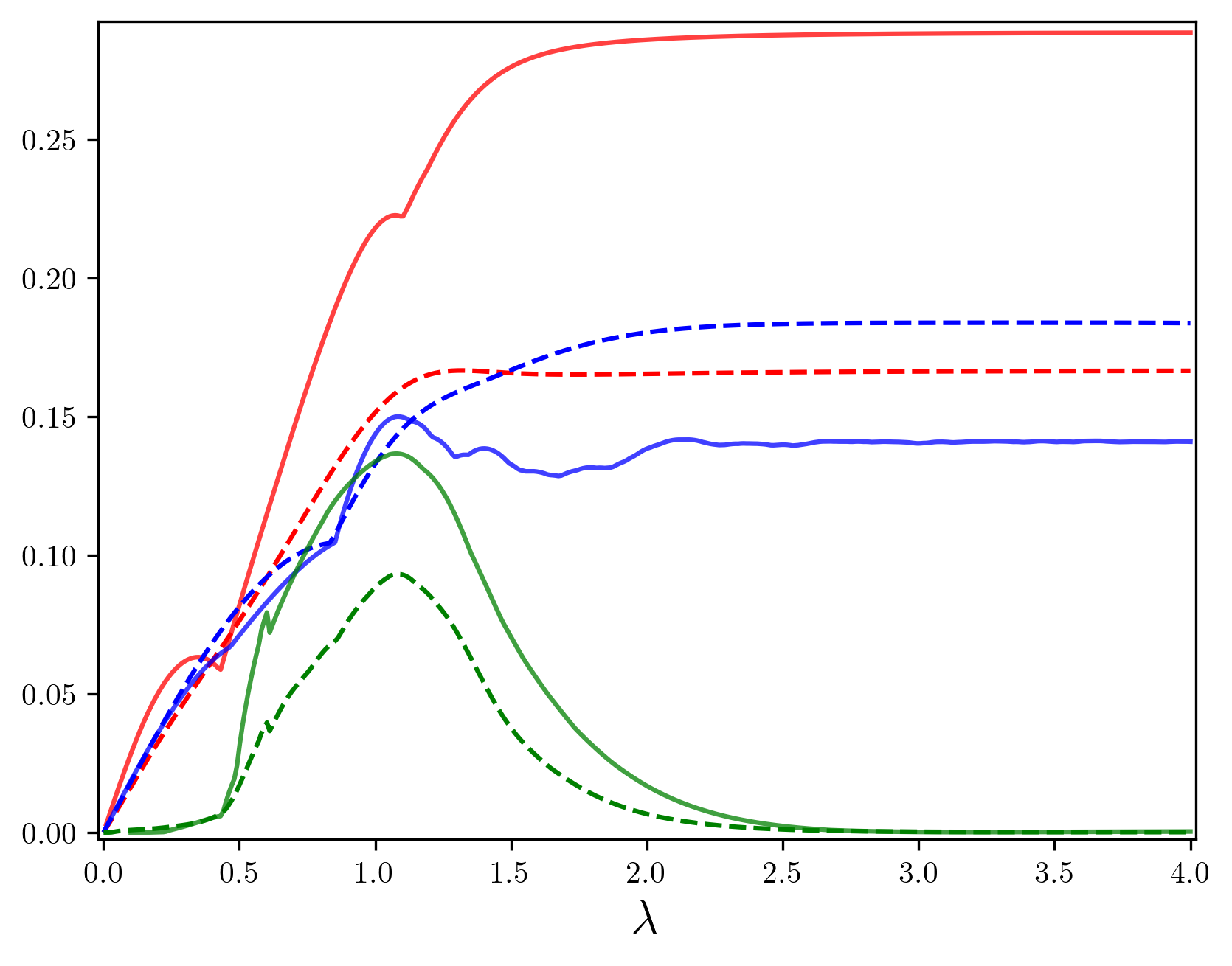}
        \caption{Numerically calculated $\sqrt{\mathcal{S}}$ (solid lines) and $\mathcal{N}$ (dashed lines) for various steady states for the symmetry-respecting bath. Red: $p_e = 2/3, p_o = 1/3,\beta_e = \beta_o = 90$. Blue: $p_e = p_o = 0.5$, $\beta_e = 90, \beta_o = 1$ (blue). Green: same as blue, but with $\varepsilon\sigma_x$ added to the closed-system Hamiltonian, with $\varepsilon = 0.1\Delta$. For all calculations, $\Delta = 2$, and for Redfield calculations, we use $\gamma_e = \gamma_o = 10^{-5}$ and assume that the even and odd baths both have spectral density $J(\omega) = \omega \exp(-\omega/\Delta)$ (detailed in Supplemental Material).}
        \label{Sym_Bath_Cross_Section}
\end{figure}

If the even (odd) bath is held at inverse temperature $\beta_e$ ($\beta_o$), then since there are no transitions between the even and odd subspaces, these subspaces thermalize at inverse temperatures $\beta_e$ and $\beta_o$ respectively, with the total populations $p_e$ and $p_o$ of each subspace unchanged from their initial values. Through the CSV condition, it is clear that negativity can be enhanced by tuning the populations $p_e$ and $p_o$. For instance, if there is more population in the even subspace than in the odd one, there will be less odd population to support the larger even-even 
 coherences, partial transposed into odd-odd ones. This leads to a large buildup of $\mathcal{S}$ and $\mathcal{N}$, even in the strong-coupling regime, with a larger difference $p_e - p_o$ yielding a larger negativity. This is corroborated numerically in Fig. \ref{Sym_Bath_Cross_Section}, where we show in red the numerical $\mathcal{N}$ and $\sqrt{\mathcal{S}}$ of the steady state when $p_e = 2/3, p_o = 1/3$, with $\beta_e = \beta_o = 90$. Unlike in the original QRM, $\mathcal{N}$ and $\mathcal{\sqrt{S}}$ both remain large even in the strong-coupling regime - the large entanglement that one would intuitively expect to accompany strong coupling is manifest in the QRM if one can selectively populate one of the parity subspaces.

 Alternatively, the CSV condition also suggests that negativity can be enhanced via a temperature difference. For example, suppose that the temperature of the odd bath is increased, while the even bath is kept at low temperature. For large $\lambda/\Delta$, the low-temperature thermal state of the even subspace has large even-even coherences, while the high-temperature odd subspace thermal state has its population spread across many excited states, so that any individual state has a low population. These depopulated odd states can no longer support the partial transposed even-even coherences, leading to a larger $\mathcal{S}$, and therefore a larger $\mathcal{N}$, in the strong-coupling regime. This is confirmed numerically in blue in Fig. \ref{Sym_Bath_Cross_Section}, which shows the steady state $\mathcal{N}$ and $\sqrt{\mathcal{S}}$ when $p_e = p_o = 0.5$, $\beta_e = 90, \beta_o = 1$. As expected, $\mathcal{S}$ remains non-negligible in the strong-coupling regime, leading to an increased $\mathcal{N}$ compared to the original QRM. 

Remarkably, this means that a higher temperature actually leads to \textit{larger} negativity in the strong-coupling regime, despite the general expectation that high temperatures destroy entanglement. Given that this surprising result was rationalized using the connection between the CSV condition and $\mathbb{Z}_2$ symmetry, it is not surprising that it only occurs in the presence of the symmetry. Indeed, if one introduces the symmetry-breaking term $\varepsilon \sigma_x$ into the closed-system Hamiltonian, while retaining the parity-respecting system-bath interaction, one finds that the resulting non-equilibrium steady state (NESS) no longer displays enhanced negativity in the strong-coupling regime. This is seen numerically in green in Fig. \ref{Sym_Bath_Cross_Section}, which shows $\mathcal{N}$ and $\sqrt{\mathcal{S}}$ for the NESS calculated using Redfield theory \cite{pollard_solution_1994,nitzan_chemical_2014,liao_quantum_2011, tscherbul_quantum_2015} when $\varepsilon = 0.1\Delta$ and $\beta_e = 90, \beta_o = 1$. Not only is the increased $\mathcal{S}$ and $\mathcal{N}$ in the strong-coupling regime lost when comparing to the $\varepsilon = 0$ case, but also when compared to the green curve in Fig. \ref{JC_Rabi_Cross_Section}, one sees that the negativity in the other regimes is significantly weakened compared to the thermal state of the QRM with $\varepsilon = 0.1\Delta$. This corresponds to the usual expectation that increasing the temperature should inhibit entanglement. Once again, we see that the presence of the $\mathbb{Z}_2$ symmetry is critical to the negativity properties of the QRM, and the CSV condition illuminates the reason for this.

\textit{Conclusion}--- We have proposed the use of the ``Cauchy-Schwarz Violation" $\mathcal{S}(\rho) = \sum_{i > j} \max\{|\rho_{ij}^{PT}|^2 - \rho_{ii}^{PT}\rho_{jj}^{PT}, 0 \}$ to probe the entanglement properties of mixed states in open quantum systems. Although nonzero $\mathcal{S}$ is only a sufficient condition for nonzero of negativity, analytical and numerical examples demonstrate that in fact $\mathcal{S}$ has a close  correspondence with the negativity, making it a useful tool for predicting the entanglement properties of mixed states. The ``local" nature of $\mathcal{S}$ makes it more analytically tractable than the inherently ``global" negativity and also makes $\mathcal{S}$ a powerful numerical tool for assessing entanglement in large systems where determining the negativity can be computationally difficult. This local character also gives $\mathcal{S}$ a physically intuitive meaning: it represents a local imbalance of population and coherence. The simple interpretation of $\mathcal{S}$ allows for intuitive understanding of entanglement behavior and enables connections between entanglement and other physical properties of a system such as symmetries. This was exemplified by the symmetry arguments for the contrasting negativity in the JCM and QRM, as well as the successful predictions of the impact of further modifications to the QRM on its symmetry.

While the focus of this work has been on the (non-)equilibrium steady states of simple models, the general principles should extend both to the time-dependent domain and to more realistic systems. In time-evolving open systems, $\mathcal{S}$ should provide a direct way to understand how the combination of population transport and the creation and destruction of coherences due to bath interactions lead to the dynamics of negativity. Furthermore, the close connection between symmetry and entanglement revealed by $\mathcal{S}$ provides a way to extend studies of model systems to more realistic settings: even if a real system has different microscopic interactions than a model, it should still show similar negativity so long as it obeys the same symmetries. On the other hand, one must be careful not to use models with more symmetries than the physical system of interest, because extra symmetries can drastically alter negativity properties, as seen from comparing the results for the JCM and QRM in this work.

Finally, in this work the CSV condition and its application to the JCM and QRM have been cast in the terms of the partial transposed density operator $\rho^{PT}$, consistent with the literature on negativity. While this language is useful for demonstrating the practical implementation of the CSV approach, for further physical interpretation it is insightful to reexpress the CSV condition in terms of the coherences and populations of the original density operator $\rho$, via $\mathcal{S}(\rho) = \sum_{i > k, j > l}|\langle k,j|\rho|i,l\rangle|^2 - \langle i,j|\rho|i,j\rangle \langle k,l|\rho|k,l\rangle$. This perspective and the additional insights it provides about the connection between coherence and entanglement are the subject of ongoing research.







\begin{acknowledgements}
\textit{Acknowledgements}---This material is based upon US Air Force Office of Scientific Research (AFOSR) under grant FA9550-20-1-0354 and support from the Natural Sciences and Engineering Research Council of Canada (NSERC) and Centre for Quantum Information and Quantum Control (CQIQC). 
\end{acknowledgements}
\appendix
\section*{Supplemental Material}
\textit{Analytical Results for JCM}---Here we give more details about the analytical results for the JCM that are used in the main text. Because $H_{JCM}$ conserves the excitation number $N = (\sigma_z + 1) + a^\dagger a$ and each of the excitation number subspace is no larger than two-dimensional, it is straightforward to find the energies and eigenstates of $H_{JCM}$
 \cite{fan_quantum_2020}.  Let $|\uparrow, m \rangle$ ($|\downarrow, m\rangle$) denote the state with the TLS in the up (down) state and with $m$ bosonic excitations. Then, for a given excitation number $n \ge 1$, let $|\varepsilon_{n,+}\rangle$ ($|\varepsilon_{n,-}\rangle$) denote the energy eigenstate of higher (lower) energy with this excitation number, and label the associated energy as $E_{n, +}$ ($E_{n,-}$); the single state with zero excitations is labeled as $|\varepsilon_{0,0}\rangle$ with associated energy $E_{0,0}$. A straightforward calculation shows that the energies and eigenstates of $H_{JCM}$ are:  
\begin{equation}
\begin{split}
\label{JCM_Eigenstates}
|\epsilon_{0,0} \rangle &= |\downarrow, 0\rangle, \;\;\; E_{0,0} = -\Delta/2 \\
|\epsilon_{n,+} \rangle &= \cos(\frac{\theta_n}{2})|\uparrow, n-1\rangle +  \sin(\frac{\theta_n}{2})|\downarrow,n\rangle, \\  E_{n,+} &= (n - \frac{1}{2}) + \Omega_n/2, \\
|\epsilon_{n,-} \rangle &= -\sin(\frac{\theta_n}{2})|\uparrow, n-1\rangle +  \cos(\frac{\theta_n}{2})|\downarrow,n\rangle \\ E_{n,-} &= (n - \frac{1}{2}) - \Omega_n/2
\end{split}
\end{equation}
where $\theta_n = \arctan(2\lambda\sqrt{n}/\delta)$, $\Omega_n = \sqrt{\delta^2 + 4\lambda^2 n}$ the $n$th Rabi frequency, and $\delta = \Delta - 1$ the detuning. From the expressions for the energies, one finds that $E_{0,0}$ and $E_{1,-}$ are equal when $|\lambda| = \lambda_0 = \sqrt{\Delta}$, and $E_{n,-}$ and $E_{n + 1,-}$ are equal when $|\lambda| = \lambda_n = \sqrt{2n + 1 + \sqrt{(2n + 1)^2 + \delta^2}}$. Since $\lambda_n > \lambda_{n-1}$ for all $n$, the JCM has a degenerate ground state if and only if $|\lambda| = \lambda_n$ for some $n$. Consequently, in the low-temperature limit, the thermal state approaches $\rho_{Th} = |\varepsilon_{n, -}\rangle \langle \varepsilon_{n, -}|$ when $\lambda_{n -1} < |\lambda| < \lambda_n$, and $\rho_{Th} = \frac{1}{2}(|\varepsilon_{n, -}\rangle\langle\varepsilon_{n, -}| + |\varepsilon_{n + 1, -}\rangle\langle\varepsilon_{n + 1, -}|)$ when $|\lambda| = \lambda_n$ (in both of these expressions, we understand $|\varepsilon_{n,-}\rangle$ to be $|\varepsilon_{0,0}\rangle$ when $n = 0$). Using the expressions (\ref{JCM_Eigenstates}) for the eigenstates, one can write $\rho_{Th}^{PT}$ in the product basis and find that it decomposes into $2 \times 2$ blocks, so that $\mathcal{S}$ and $\mathcal{N}$ can both be readily calculated to give:
\begin{equation}
    \mathcal{S}(\rho_{Th}) = \begin{cases} \frac{\lambda^2n}{(\Delta - 1)^2 + 4\lambda^2 n} & \lambda_{n-1} < |\lambda| < \lambda_{n} \\ \frac{1}{4}\left(\frac{\lambda^2n}{(\Delta - 1)^2 + 4\lambda^2 n} + \frac{\lambda^2(n+1)}{(\Delta - 1)^2 + 4\lambda^2 (n+1)}\right) & |\lambda| = \lambda_n
    \end{cases}
\end{equation}
\begin{equation}
    \mathcal{N}(\rho_{Th}) = \begin{cases} \sqrt{\mathcal{S}(\rho_{Th})} &\lambda_{n-1} < |\lambda| < \lambda_{n}  \\\frac{1}{4}\left(\sqrt{\sin^4(\theta_{n+1}/2) + \sin^2(\theta_n)} - \sin^2(\theta_{n+1}/2) + \sqrt{\cos^4(\theta_n/2) + \sin^2(\theta_{n+1})}- \cos^2(\theta_n/2)\right) & |\lambda| = \lambda_n 
    \end{cases}
\end{equation}
In particular, from these expressions one can check that $\mathcal{S}$ is bounded above by $1/4$ when $|\lambda| \neq \lambda_n$ and by $1/8$ when $|\lambda| = \lambda_n$, while $\mathcal{N}$ is bounded above by $1/2$ when $|\lambda| \neq \lambda_n$ and by $1/3$ when $|\lambda| = \lambda_n$.

\textit{Computational Details and Redfield Theory}--- The JCM and QRM were numerically implemented by truncating the bosonic degree of freedom at 45 excitations, which was found to be sufficient for convergence for the parameters considered in this work. To calculate $\mathcal{S}$ and $\mathcal{N}$ in Fig. \ref{JC_Rabi_Cross_Section}, the thermal states $\rho_{Th} = \exp(-\beta H)/Tr(e^{-\beta H})$ were obtained by numerically performing matrix exponentiation. When the baths respect the QRM $\mathbb{Z}_2$ symmetry and the even and odd subspaces thermalize separately, the steady state is given by $\rho_{ss} = p_eP_e e^{-\beta_e H} P_e/\Tr(P_e e^{-\beta_e H} P_e)  + p_oP_o e^{-\beta_o H} P_o/\Tr(P_o e^{-\beta_o H} P_o)$, which can be computed by numerical matrix exponentiation to yield the red and blue curves in Fig. \ref{Sym_Bath_Cross_Section}. Finally, to determine the steady state when the symmetry-broken term is introduced, we model the time-evolution of the density operator using the Redfield formalism. In the energy eigenbasis of the closed-system Hamiltonian, the time evolution is \cite{nitzan_chemical_2014}:
\begin{equation}
\label{Redfield_EOM}
    \dot{\rho}_{mn} = -i\omega_{mn} + \sum_{o,p}\mathcal{R}_{mn,op}\sigma_{op}
\end{equation}
with $\omega_{mn}$ the energy gap between states $m$ and $n$, and:
\begin{equation}
    \mathcal{R}_{mn,op} = \Gamma_{pn,mo} + \Gamma_{om, np}^* - \left(\delta_{np} \sum_q \Gamma_{mq,qo}\right) - \left(\delta_{mo} \sum_{q} \Gamma_{nq,qp}^*\right)
\end{equation} where, assuming the environment consists of independent bosonic baths $j$ with spectral densities $J^j$ and coupling to the system via some system operator $T^j$, these $\Gamma_{mn,op}$ can be written as \cite{pollard_solution_1994}:
\begin{equation}
    \Gamma_{mn,op} = \begin{cases} \pi \sum_jT^j_{mn}T^j_{op}J^j(\omega_{op}) \frac{1}{e^{\beta \omega_{op}} - 1} & \omega_{op} > 0 \\
    \pi \sum_jT^j_{mn}T^j_{op}J^j(-\omega_{op}) \frac{e^{-\beta \omega_{op}}}{e^{-\beta \omega_{op}} - 1} & \omega_{op} < 0 \\
    0 & \omega_{op} = 0 \end{cases}
\end{equation}
with the sum taken over the baths $j$. For the putative symmetry-respecting system-bath interaction introduced in the main text, we have two baths $j = e$ (the even bath) and $j = o$ (the odd bath); we set $T^e = \gamma_eP_e\sigma_x(a^\dagger + a)P_e$ and $T^o = \gamma_oP_o\sigma_x(a^\dagger + a)P_o$, and for both the even and the odd baths we assume an Ohmic spectral density $J^j(\omega) = \omega \exp(-\omega/\Delta)$. With these substitutions, the steady state can be obtained by numerically solving for the kernel of the time-translation generator $\mathcal{L}$, defined by $\mathcal{L}\rho = \dot{\rho}$ with $\dot{\rho}$ given by Eq. (\ref{Redfield_EOM}).

\begin{figure}
    \centering
\begin{subfigure}{0.45\linewidth}
\includegraphics[width = \linewidth]{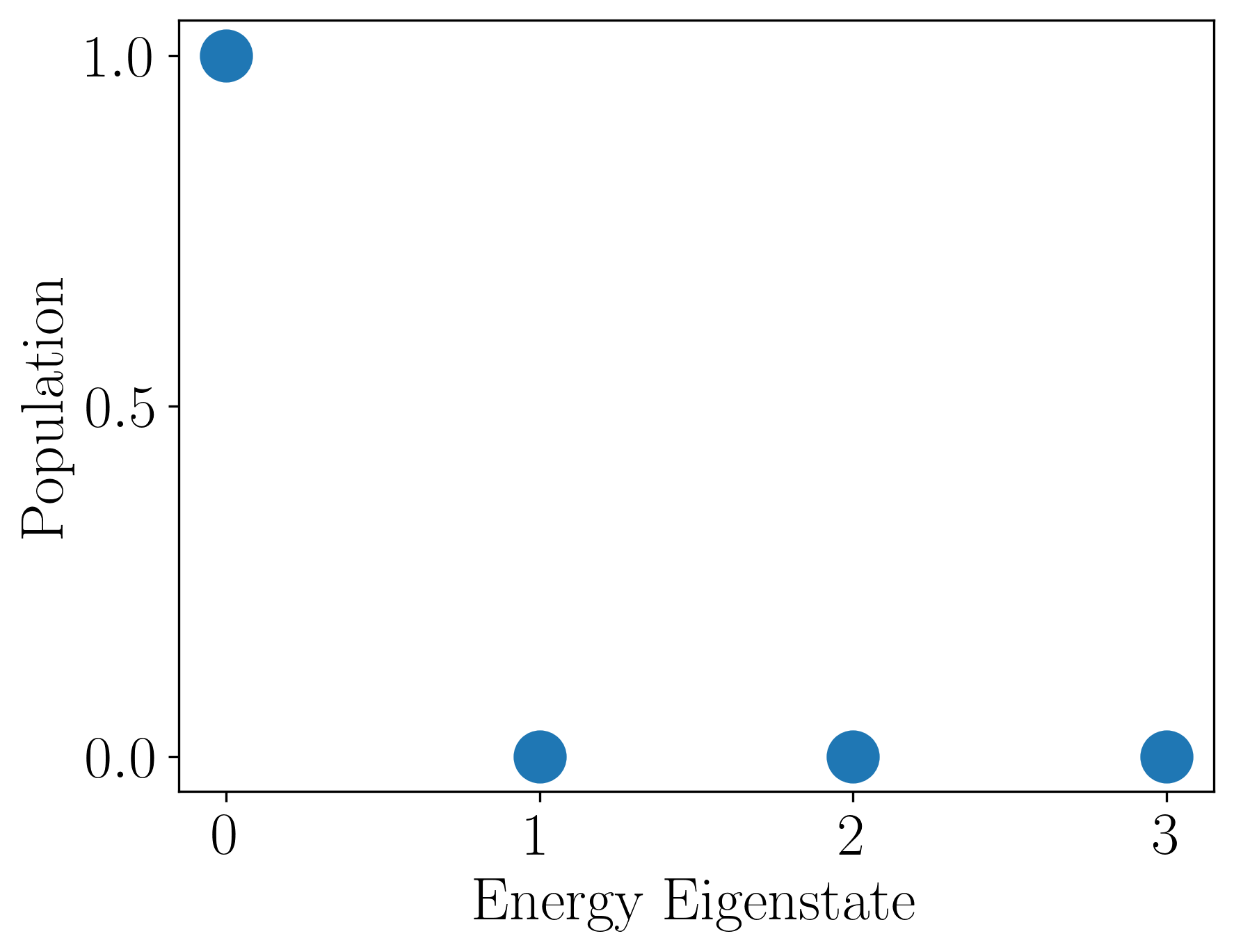}
\caption{}
\label{3a}
\end{subfigure}
\hspace{0.5cm}
\begin{subfigure}{0.45\linewidth}
\includegraphics[width = \linewidth]{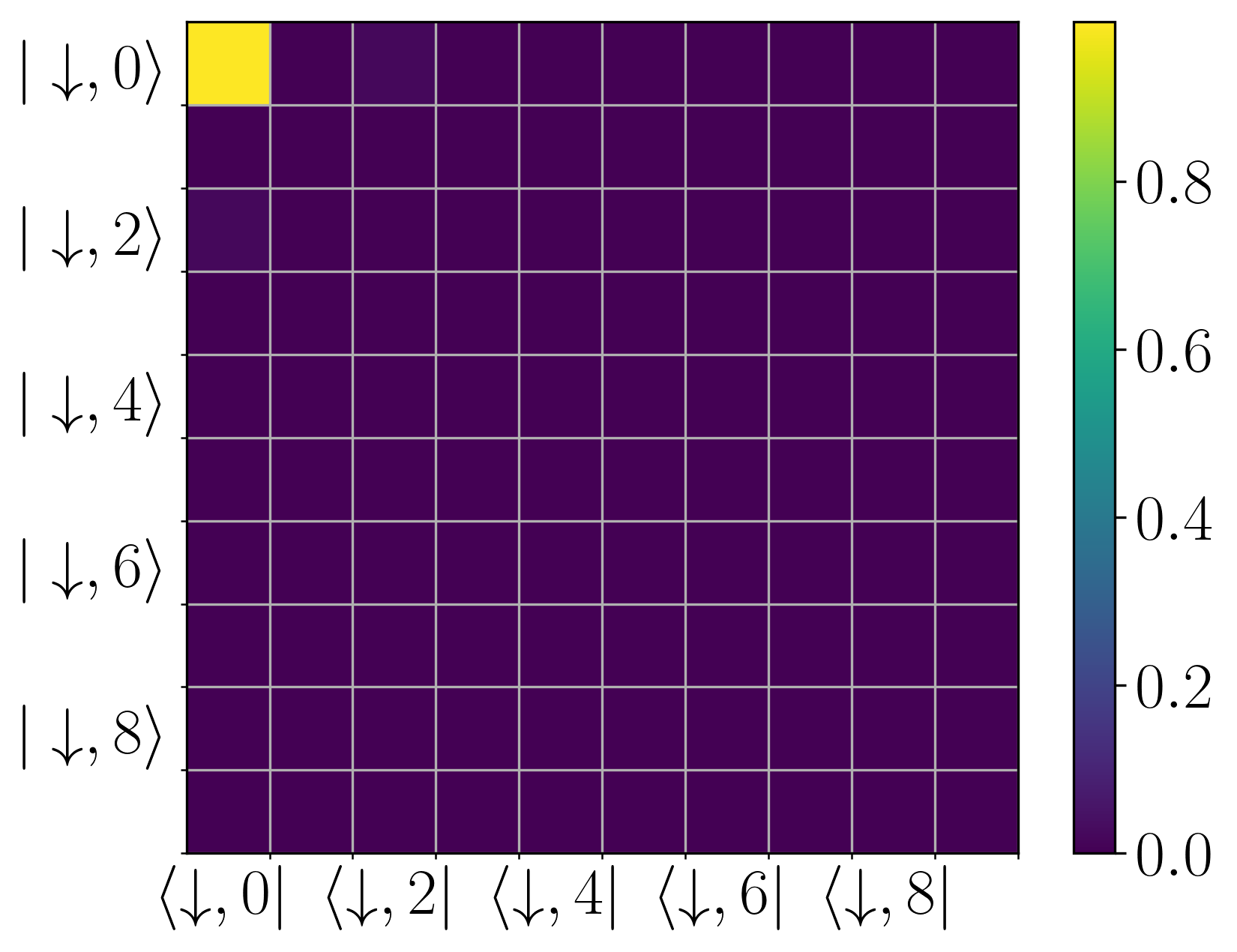}
\caption{}
\label{3b}
\end{subfigure}
\begin{subfigure}{0.45\linewidth}
\includegraphics[width = \linewidth]{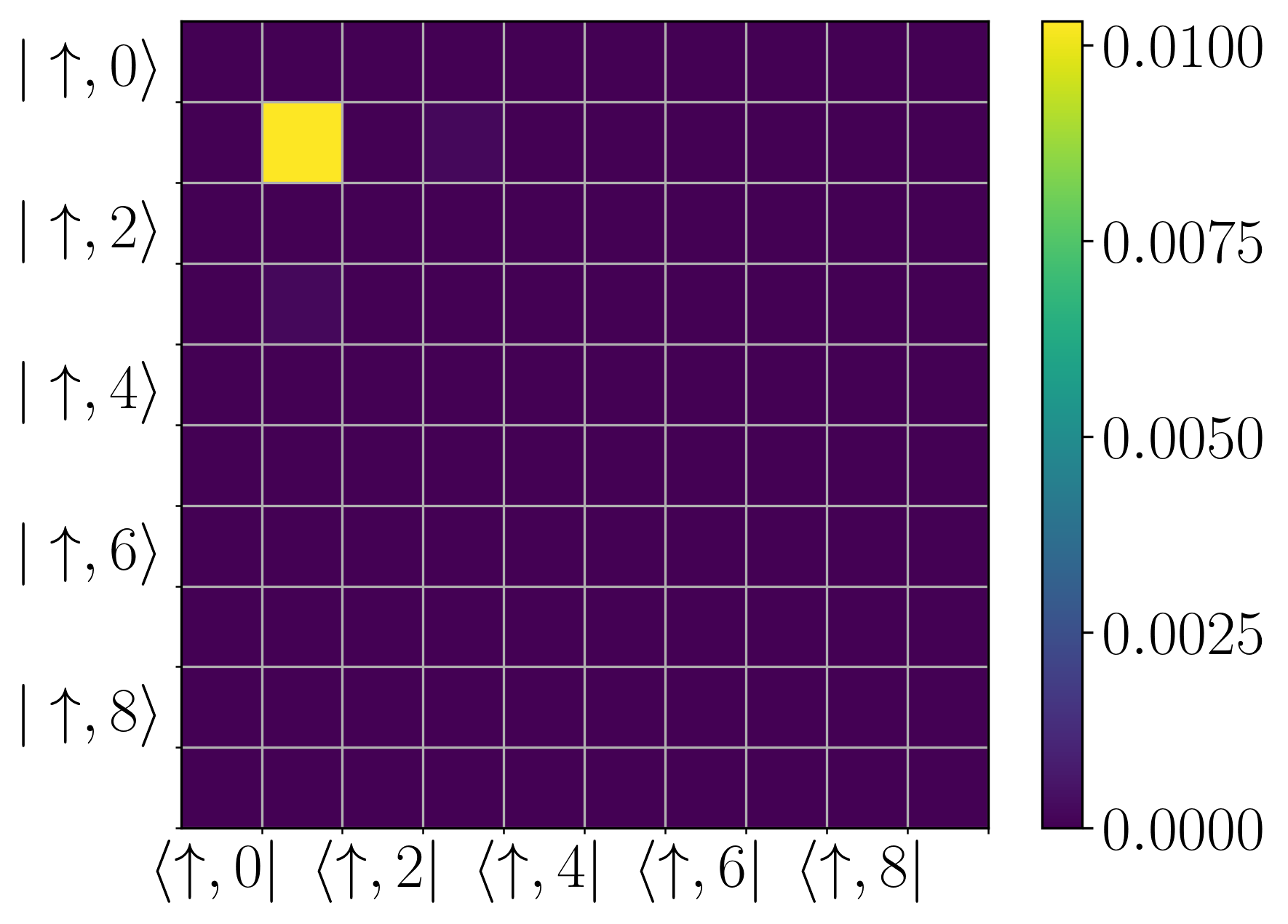}
\caption{}
\label{3c}
\end{subfigure}
\hspace{0.5cm}
\begin{subfigure}{0.45\linewidth}
\includegraphics[width = \linewidth]{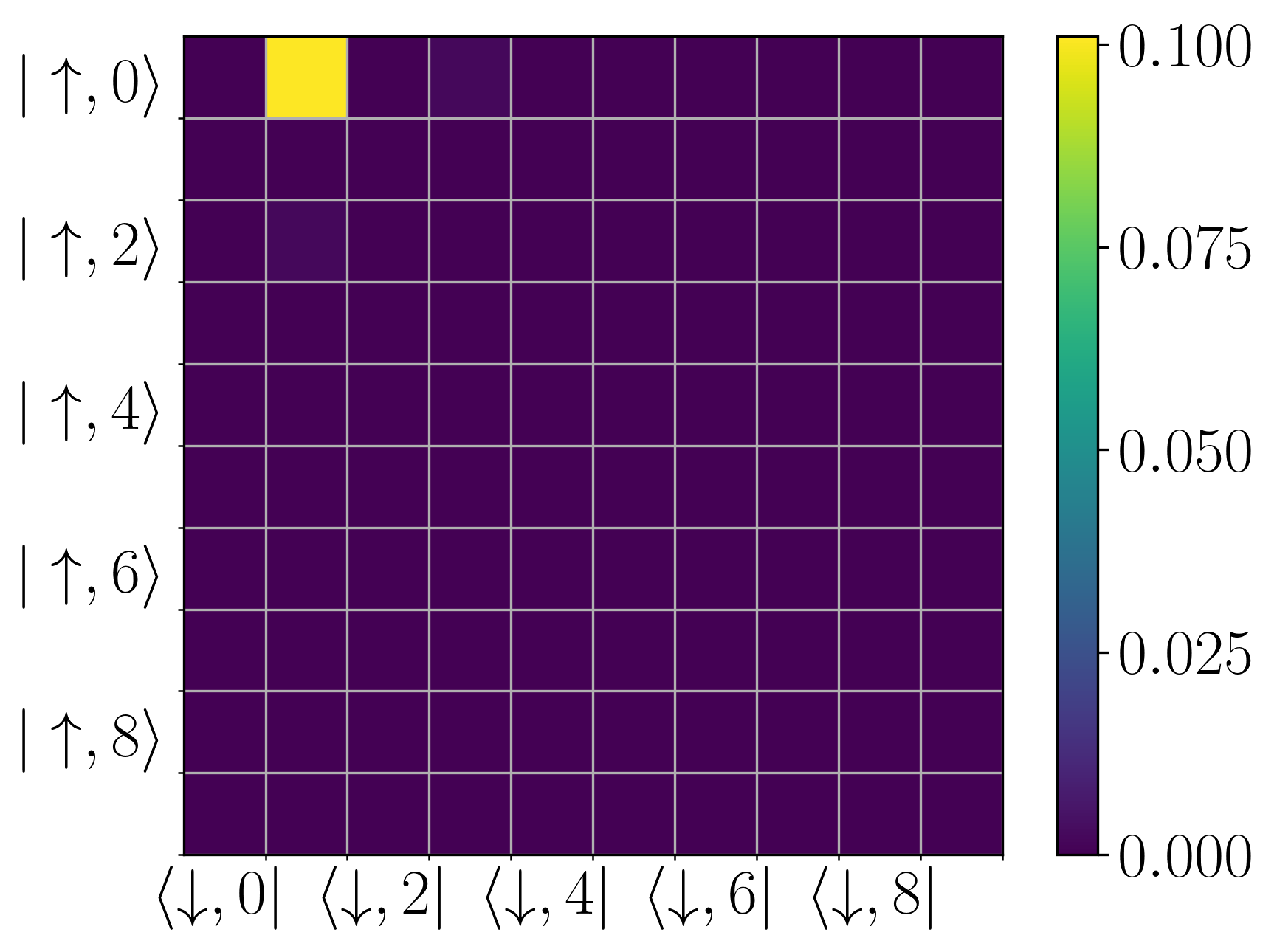}
\caption{}
\label{3d}
\end{subfigure}
        \caption{Numerical results for the thermal state of the QRM for $\lambda = 0.3,\, \Delta = 2, \,\beta = 90$. (a): Numerically calculated thermal energy eigenstate populations. (b), (c), (d): Numerically calculated magnitudes of certain matrix entries of the thermal density operator, with each grid square corresponding to one matrix entry as labeled by the axes.}
        \label{small_lambda_dens_op}
\end{figure}

\begin{figure}
    \centering
\begin{subfigure}{0.45\linewidth}
\includegraphics[width = \linewidth]{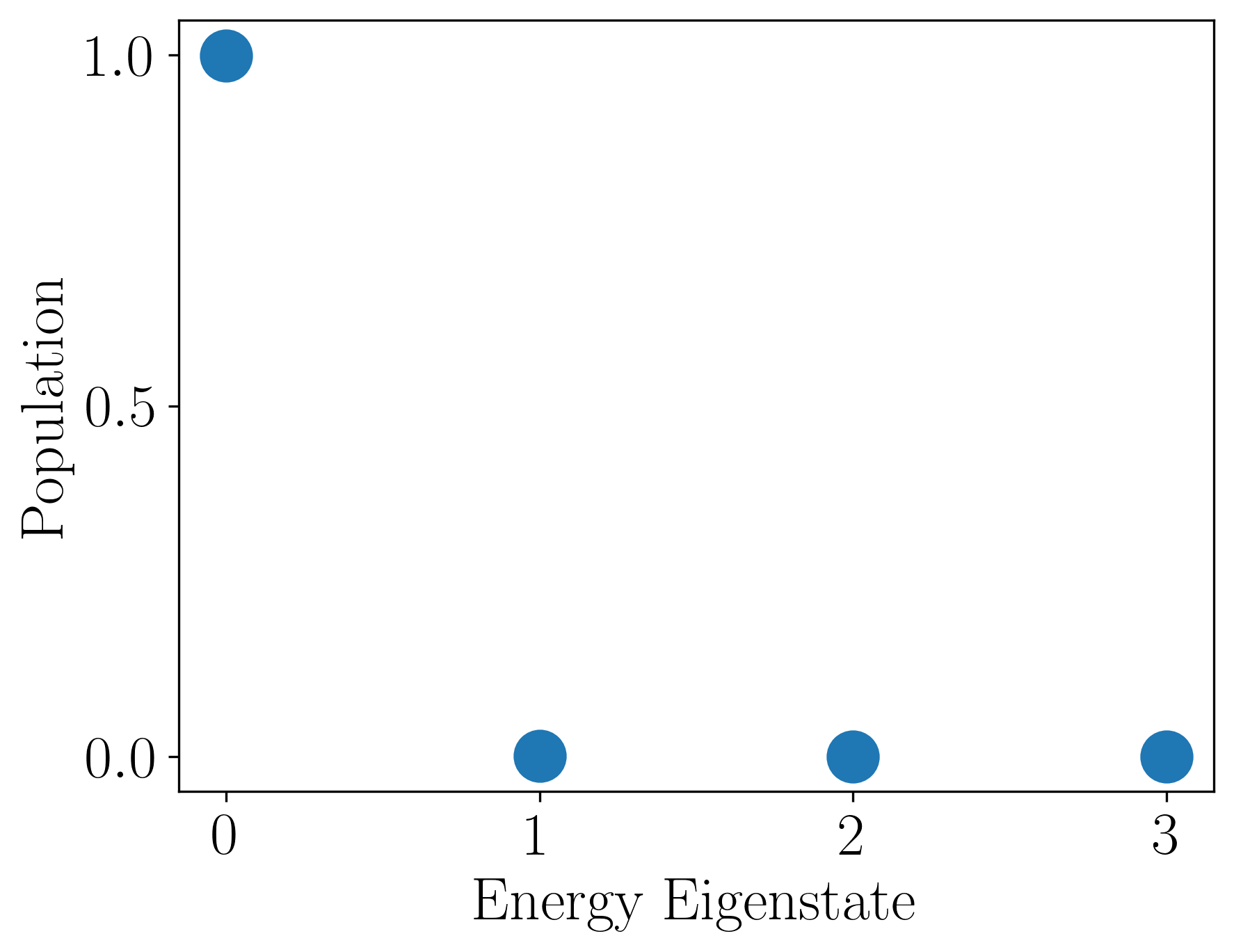}
\caption{}
\label{4a}
\end{subfigure}
\hspace{0.5cm}
\begin{subfigure}{0.45\linewidth}
\includegraphics[width = \linewidth]{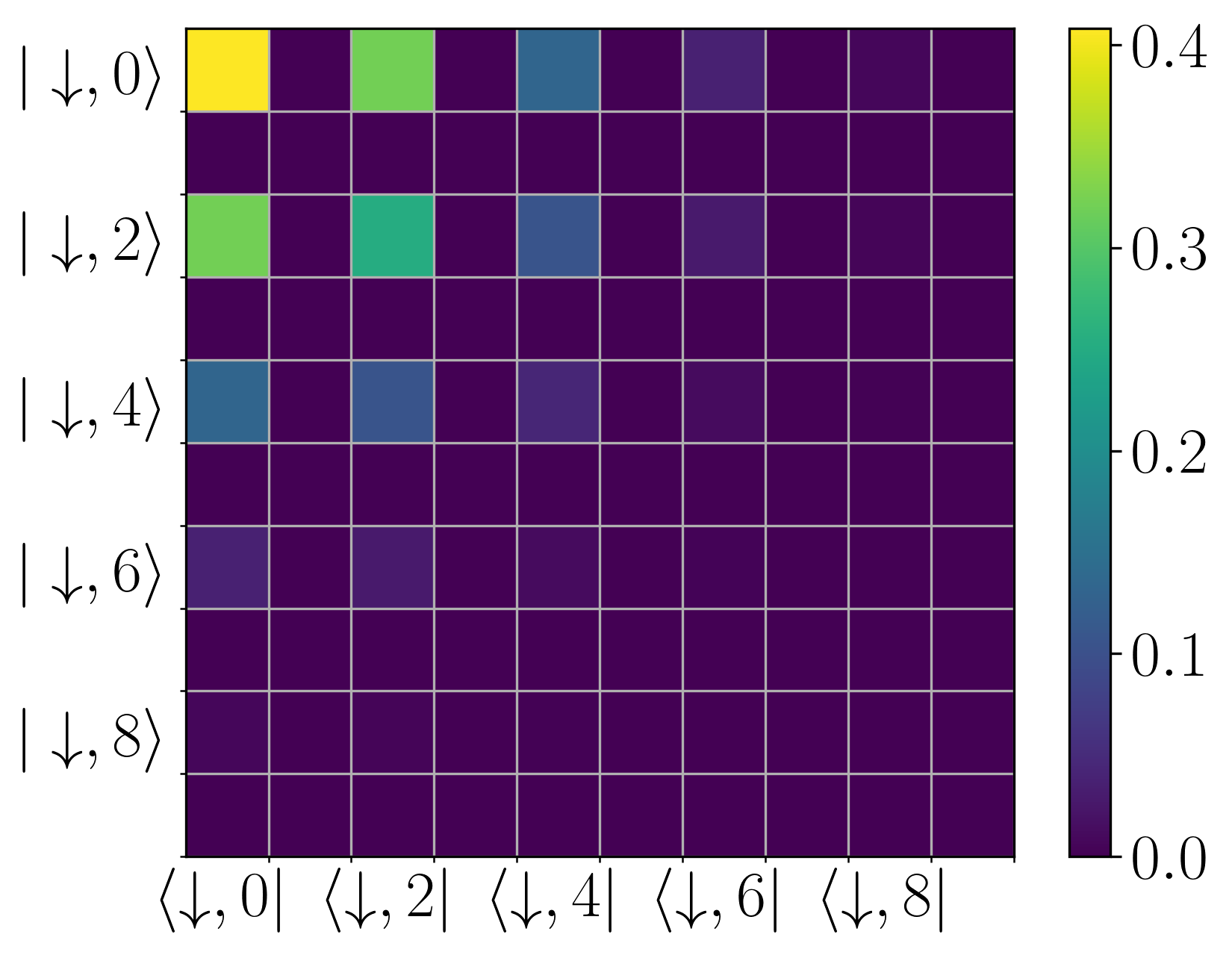}
\caption{}
\label{4b}
\end{subfigure}
\begin{subfigure}{0.45\linewidth}
\includegraphics[width = \linewidth]{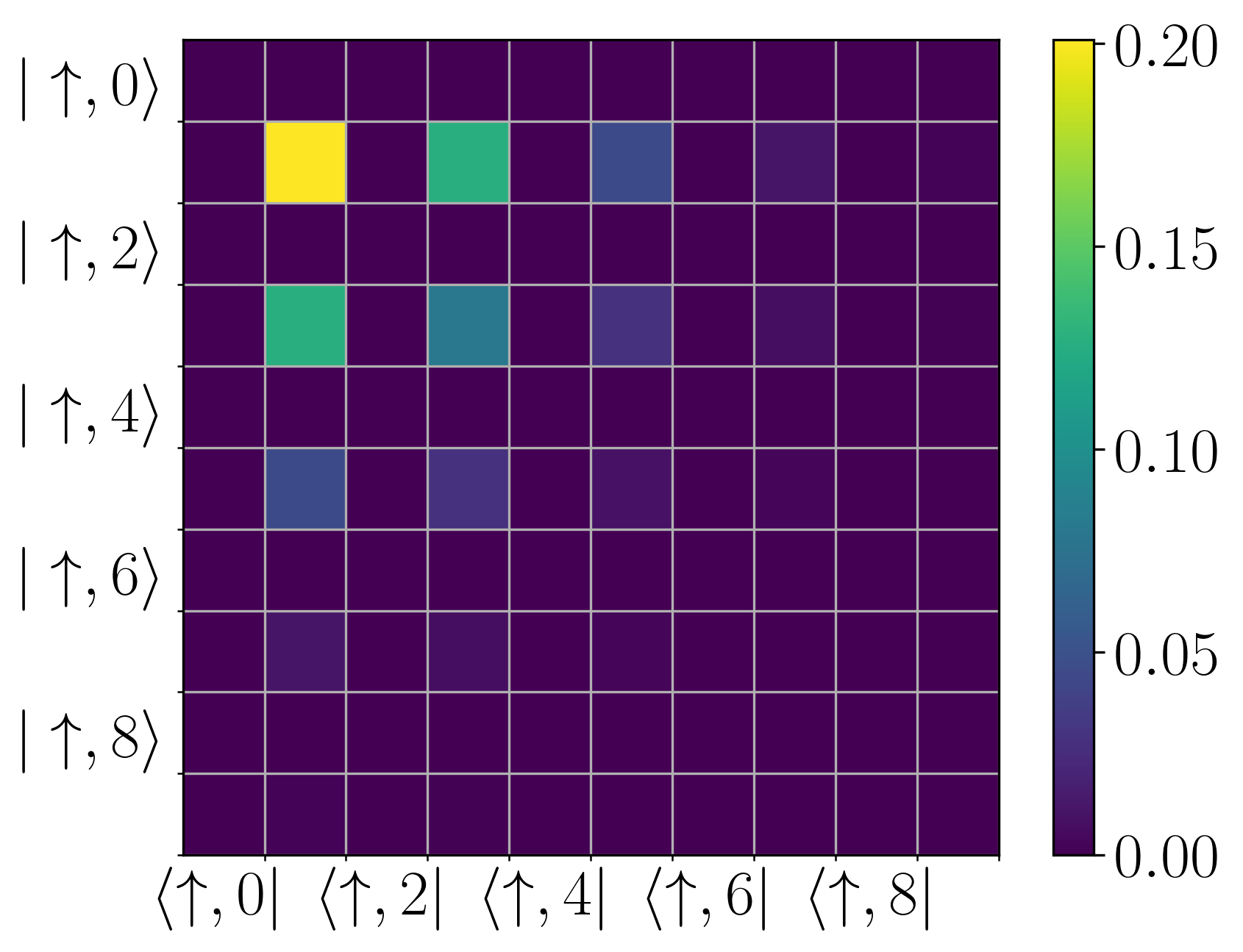}
\caption{}
\label{4c}
\end{subfigure}
\hspace{0.5cm}
\begin{subfigure}{0.45\linewidth}
\includegraphics[width = \linewidth]{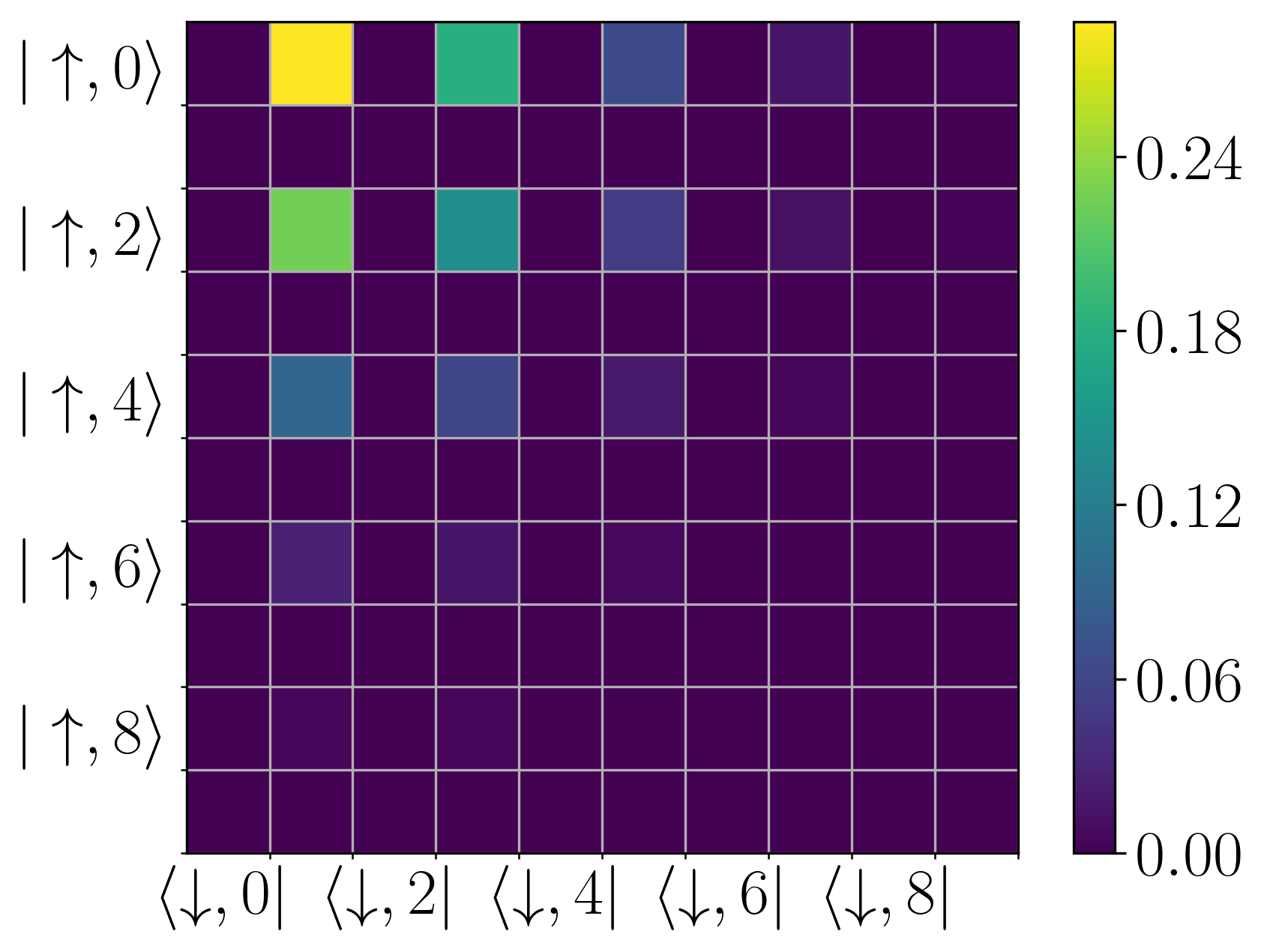}
\caption{}
\label{4d}
\end{subfigure}
 \caption{Same as Fig. \ref{small_lambda_dens_op}, but now for $\lambda = 1.3$.}    \label{medium_lambda_dens_op}
\end{figure}

\begin{figure}
    \centering
\begin{subfigure}{0.45\linewidth}
\includegraphics[width = \linewidth]{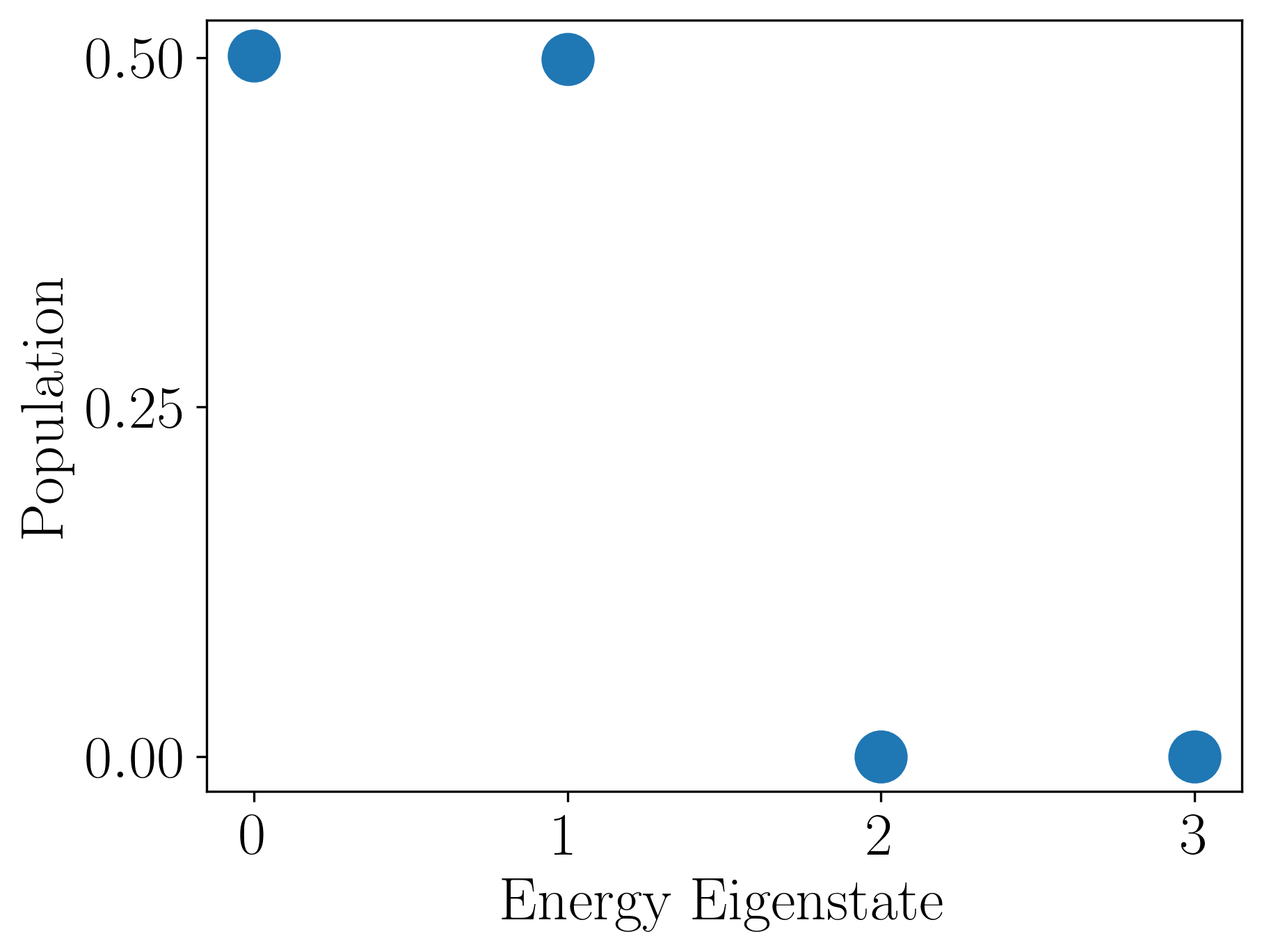}
\caption{}
\label{5a}
\end{subfigure}
\hspace{0.5cm}
\begin{subfigure}{0.45\linewidth}
\includegraphics[width = \linewidth]{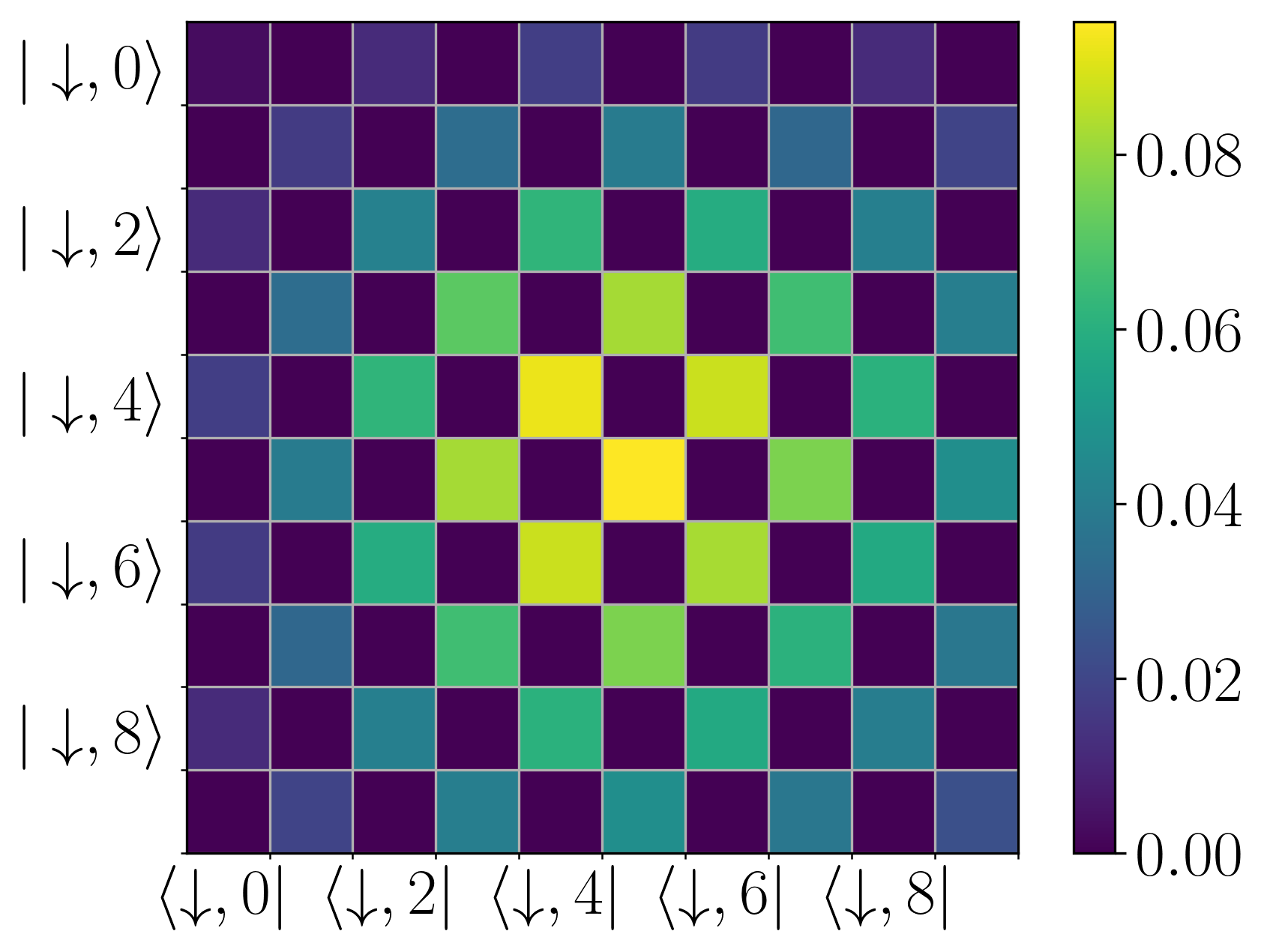}
\caption{}
\label{5b}
\end{subfigure}
\begin{subfigure}{0.45\linewidth}
\includegraphics[width = \linewidth]{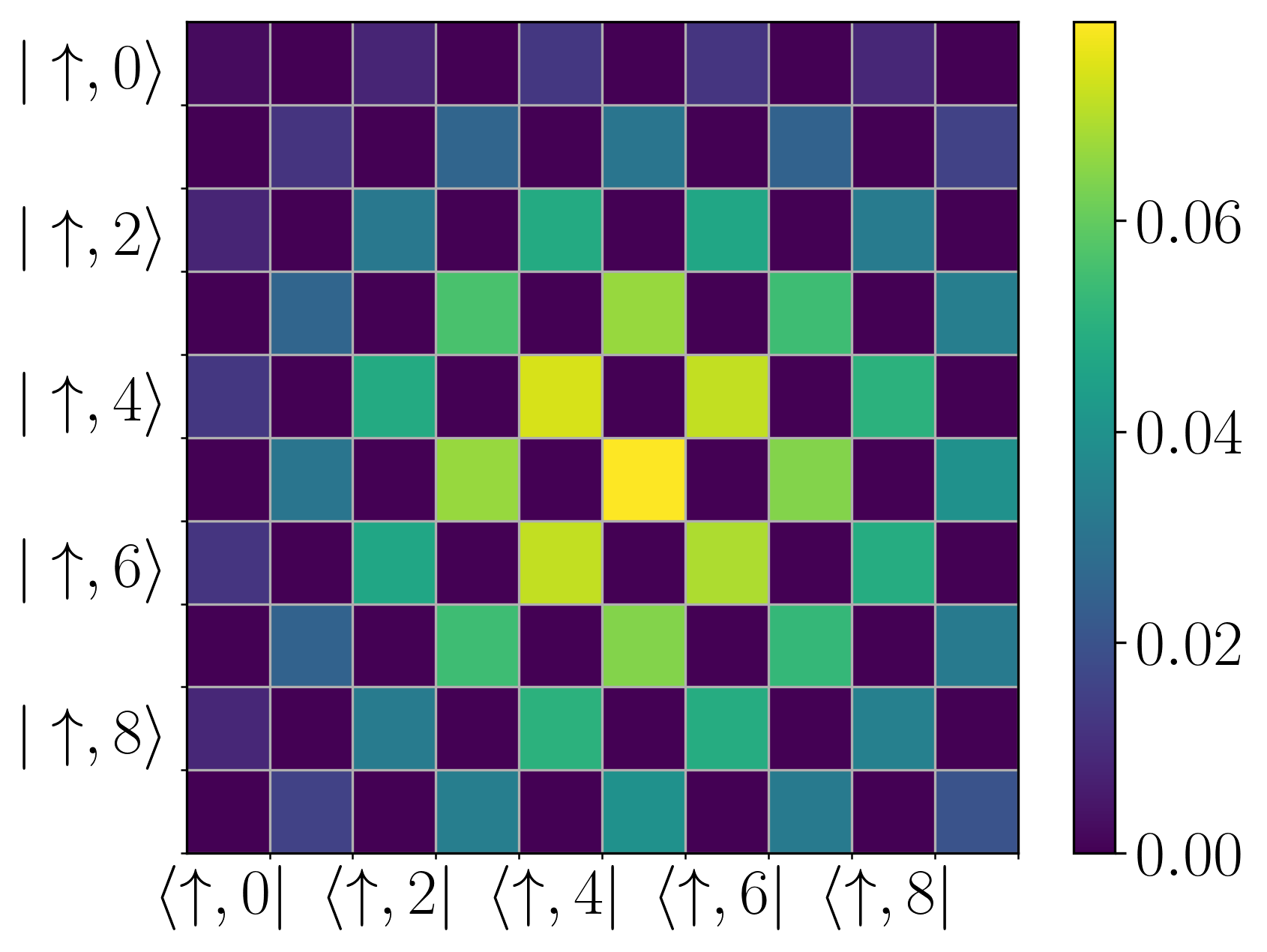}
\caption{}
\label{5c}
\end{subfigure}
\hspace{0.5cm}
\begin{subfigure}{0.45\linewidth}
\includegraphics[width = \linewidth]{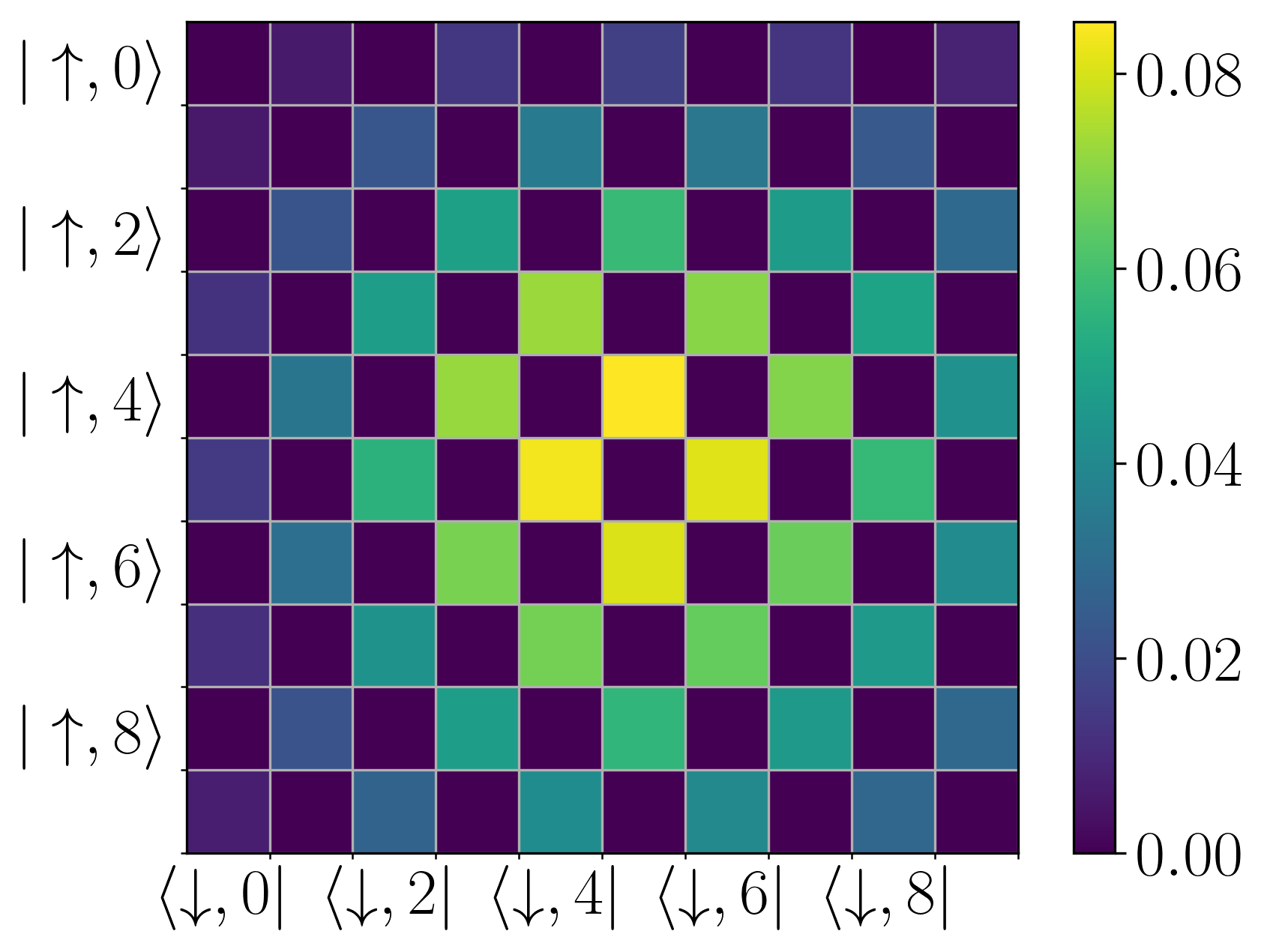}
\caption{}
\label{5d}
\end{subfigure}
        \caption{Same as Fig. \ref{small_lambda_dens_op}, but now for $\lambda = 2.3$.}
        \label{large_lambda_dens_op}
\end{figure}

\textit{Density Operator Plots for QRM}---As further evidence for the CSV arguments presented in the main text, we examine the structure of the thermal density operators obtained from the numerics. As an example, in Figs. \ref{small_lambda_dens_op}-\ref{large_lambda_dens_op}, we show the thermal density operators for $\lambda  =0.3$, $\lambda = 1.3$ and $\lambda = 2.3$, with $\Delta = 2$ and $\beta = 90$ held fixed. For $\lambda  = 0.3$ (the weak coupling regime), a plot of the thermal energy eigenstate populations in Fig. \ref{3a} shows that only the lowest energy eigenstate has non-negligible population in equilibrium; since this eigenstate is in the even subspace, it results in the product basis having only a significant population in the even states $|\downarrow, 0\rangle$ and $|\uparrow, 1\rangle$ (Fig. \ref{3b}-\ref{3c}). The even-even coherence arising between these states is seen in Fig. \ref{3d}; when this becomes an odd-odd coherence upon partial tranposition, there are no odd populations able to support it, leading to a nonzero, but small, $\mathcal{S} = 0.0102$ (numerically calculated). This corresponds to a small, nonzero numerically calculated negativity of $\mathcal{N} = 0.101$. These observations agree with the general discussion of the QRM in the small-coupling regime in the main text.

For $\lambda = 1.3$ (the intermediate coupling regime), the ground state is still the only significantly populated state (Fig. \ref{4a}) in equilibrium, but it is distributed over several product basis states due to the stronger coupling $\lambda$ (Fig. \ref{4b}-\ref{4c}). This results in many large even-even coherences (Fig. \ref{4d}) that partial transpose to give $\mathcal{S} = 0.207$ (a larger value than when $\lambda = 0.3$) due to the lack of odd population. This corresponds to a larger negativity $\mathcal{N} = 0.454$, in agreement with the discussion in the main text.

Finally, when $\lambda = 2.3$ (strong coupling regime), there is now approximately equal population of the lowest two energy eigenstates (Fig. \ref{5a}) resulting in large even and odd populations and associated even-even and odd-odd coherences (Fig. \ref{5b}-\ref{5d}). The ``checkerboard" pattern in the density operator is a result of the lack of even-odd coherences, which is a consequence of the fact that the energy eigenstates still have definite parity, and thus the presence of both even and odd populations is due to the incoherent mixture of the two lowest-energy states. The simultaneous presence of even and odd populations enables the density operator to support the partial transposed coherences, leading to a smaller $\mathcal{S} = 0.00334$ (much smaller than the $\lambda = 0.3$ or $\lambda = 1.3$ case). The negativity also shows the expected decrease from the intermediate $\lambda$ regime: now $\mathcal{N} = 0.0122$. This strong suppression of negativity in the strong-coupling regime matches the predictions from the CSV condition in the main text. 

\bibliographystyle{unsrt}
\bibliography{JCM_QRM_Intro_Citations,Redfield_citations,tls_references,Rabi_JC_references,Mathematical_Negativity_Results,spin_chain_citations, sym_bath_references}
\end{document}